% Torsional Directed Walks, Entropic Elasticity, and DNA Twist Stiffness
% preprint version
%  Moroz Nelson
%
% print this using plain TeX, harvmac, epsf
% 9 July

%SWITCHES
\long\def\cut#1{}
\long\def\optional#1{}
\def\pagin#1{}
\def\boringfonts{y}   % include for export
\def\figflag{y} \input epsf  % remove for submission

\input harvmac  %\input philmac

%%%%%%%%%%%%%%%%%%%%%%%%%%%%%%%%%%%%%%%%%%%%%%%%%%%%%%%%%%%%%%%%%%%%%%
%%%%%%%%%%%%% philmac.tex %%%%%%%%%%%%%

\def\fonttest{y}
%macros useful in addition to mac.tex
% WARNING: what was once \fig is now \pfig !!
%           "    "   "   \listtoc "  \plisttoc
%           "    "   "   \nfig       \pnfig
%           "    "   "   \del        \pdel

% first a full nine-point font set
\ifx\boringfonts\fonttest\def\ninepoint{}
\else
\font\ninerm=cmr9\font\ninei=cmmi9\font\nineit=cmti9\font\ninesy=cmsy9
\font\ninebf=cmbx9\font\ninesl=cmsl9\font\ninett=cmtt9
\def\ninepoint{\def\rm{\fam0\ninerm}% switch back to 10-point type
\textfont0=\ninerm \scriptfont0=\sevenrm \scriptscriptfont0=\fiverm
\textfont1=\ninei  \scriptfont1=\seveni  \scriptscriptfont1=\fivei
\textfont2=\ninesy \scriptfont2=\sevensy \scriptscriptfont2=\fivesy
\textfont\itfam=\nineit \def\it{\fam\itfam\nineit} \def\sl{\fam\slfam\ninesl}
\textfont\bffam=\ninebf \def\bf{\fam\bffam\ninebf}
\def\tt{\fam\ttfam\ninett}\rm}
\fi

\hyphenation{anom-aly anom-alies coun-ter-term coun-ter-terms
dif-feo-mor-phism dif-fer-en-tial super-dif-fer-en-tial dif-fer-en-tials
super-dif-fer-en-tials reparam-etrize param-etrize reparam-etriza-tion}

%\input usr1:[nelson.texutil]mac.tex

%
% Tagged sections: generates a symbol with current secno and also a
%               table of contents entry
%
% The page number in the toc may be wrong when a section contains no
% text. Try modifying \tnewsec by eliminating \let\the=0 .
%
\newwrite\tocfile\global\newcount\tocno\global\tocno=1
\ifx\bigans\answ \def\tocline#1{\hbox to 320pt{\hbox to 45pt{}#1}}
\else\def\tocline#1{\line{#1}}\fi
\def\toclead{\leaders\hbox to 1em{\hss.\hss}\hfill}
\def\tnewsec#1#2{\newsec{#2}\xdef #1{\the\secno}
\ifnum\tocno=1\immediate\openout\tocfile=toc.tmp\fi\global\advance\tocno
by1
{\let\the=0\edef\next{\write\tocfile{\medskip\tocline{\secsym\ #2\toclead\the
\count0}\smallskip}}\next}% want to expand secsym now, count0 later
}
\def\tnewsubsec#1#2{\subsec{#2}\xdef #1{\the\secno.\the\subsecno}
\ifnum\tocno=1\immediate\openout\tocfile=toc.tmp\fi\global\advance\tocno
by1
{\let\the=0\edef\next{\write\tocfile{\tocline{ \ \secsym\the\subsecno\
#2\toclead\the\count0}}}\next}
}
\def\tappendix#1#2#3{\xdef #1{#2.}\appendix{#2}{#3}
\ifnum\tocno=1\immediate\openout\tocfile=toc.tmp\fi\global\advance\tocno
by1
{\let\the=0\edef\next{\write\tocfile{\tocline{ \ #2.
#3\toclead\the\count0}}}\next}
}
%
% generate rudimentary table of contents
%

%
%
%
% manuscript control
%
% Phys Rev Letters: (those jerks)
% doublespace, put footnotes among references, doublespace refs, set flag

\gdef\prlmode{F}
\long\def\optional#1{}
\def\cmp#1{#1}         %or remove for stuffy journals
%
%   Matters of taste
%
\let\narrowequiv=\equiv
\def\equiv{\;\narrowequiv\;}

%\def\tilde{\widetilde}
     %or else other way round
\fontdimen16\tensy=2.7pt\fontdimen17\tensy=2.7pt %an experiment
 %usual dup not needed with the above
%\mathsurround=1pt %screws up \dsl!

% for the drafts this is useful: (remember not to do \listrefs)
%\def\ref#1#2{\edef\tmp{$[$\string#1$]$} \tmp\edef#1{\tmp}}

% for reduction this is useful:
% \ifx\answ\bigans ...<some equation> \else ... <same with linebreaks> \fi

%
%Greek abbreviations
\def\ga{\gamma}
\def\la{\lambda}

\def\th{\theta}

%
%Curly letters
%

\def\CO{{\cal O}}

\def\CE{{\cal E}}

\def\CZ{{\cal Z}}\def\CCZ{$\CZ$}
%
%
%   Miscellaneous
%
% Usage: \boxit{3.5}{some text}
\def\boxit#1#2{
        $$\vcenter{\vbox{\hrule\hbox{\vrule\kern3pt\vbox{\kern3pt
        \hbox to #1truein{\hsize=#1truein\vbox{#2}}\kern3pt}\kern3pt\vrule}
        \hrule}}$$
}

%\def\bivector#1{{\buildrel \leftrightarrow\over #1}}

         % |#1>
         % <#1|
 %matrix element <#1|#2|#3>

  %bold nabla
         % memo: you can't just temporarily redefine \textfont2 in the
         % middle of an eqn, hence the clumsy construction above. If
         % you wanted a lot of bold symbols, it would be better to do
         % a \newfam\bsymb etc as in philmac's \curly construction.

%

\def\lfr#1#2{{\textstyle{#1\over#2}}} %       little fraction
 %     tensor up-down

%\def\THETA#1#2#3{\vartheta \hbox{${#1\atopwithdelims[]#2}$}
%               \bigl(#3|\tau\bigr)}
%\def\THETB#1#2#3#4{\vartheta \hbox{${#1\atopwithdelims[]#2}$}
%               \bigl(#3|#4\bigr)}

             % no dot i

% split exact sequence
\def\splitexact#1#2{\mathrel{\mathop{\null{
\lower4pt\hbox{$\rightarrow$}\atop\raise4pt\hbox{$\leftarrow$}}}\limits
^{#1}_{#2}}}

%semi-direct product |><
%
%   Various delbars
%

            %delbar dagger
              %delbar dagger delbar
  %delbar n dagger delbar n
            %del-bar-sub n
           %delbar
\def\pd#1#2{{\partial #1\over\partial #2}} %partial derivative
 %partial derivative with room for tilde etc
     %d'alembertian
%
%   Other things with bars
%
          % Beltrami
 % conjugate Beltrami
  % z-bar
  % q-bar
  % tau-bar
  % u-bar
  % a-bar
      % mu-bar
%
%   Various romans for math
%

  % Im, Re
\def\ex#1{{\rm e}^{#1}}                 % exponential
\def\dd{\mskip 1.3mu{\rm d}\mskip .7mu} % exterior derivative
                   % dz dzbar
                      % trace

        % DET bundle
        % DET bundle

                % det
                % sdet

%
%   Whole words
%

\def\IM{isomorphism}

\def\ie{{\it i.e.}}

%
%   Fonts
%

%\def\cmfontflag{cm}
%\ifx\fontflag\cmfontflag\font\CAPS=cmcsc10 scaled 1200
% \ifx\answ\bigans\font\speci=bbb12 scaled 833\else\font\speci=bbb10\fi
%\else
% \font\CAPS=amcsc10 scaled 1200
%\fi
\ifx\boringfonts\fonttest
\font\blackboard=cmssbx10 \font\blackboards=cmssbx10 at 7pt  % wimpy
\font\blackboardss=cmssbx10 at 5pt
\else
\font\blackboard=msym10 \font\blackboards=msym7   % cool
\font\blackboardss=msym5
\fi
\newfam\black
\textfont\black=\blackboard
\scriptfont\black=\blackboards
\scriptscriptfont\black=\blackboardss

%\def\spec#1{\hbox{\speci #1}}
                      %historical

%
\ifx\boringfonts\fonttest
\font\gothic=cmssbx10 \font\gothics=cmssbx10 at 7pt  % wimpy substitute
\font\gothicss=cmssbx10 at 5pt
\else
\font\gothic=eufm10 \font\gothics=eufm7
\font\gothicss=eufm5
\fi
\newfam\gothi
\textfont\gothi=\gothic
\scriptfont\gothi=\gothics
\scriptscriptfont\gothi=\gothicss

{\catcode`\@=11\gdef\oldcal{\fam\tw@}}
\newfam\curly
\ifx\boringfonts\fonttest\else
\font\curlyfont=eusm10 \font\curlyfonts=eusm7
\font\curlyfontss=eusm5
\textfont\curly=\curlyfont
\scriptfont\curly=\curlyfonts
\scriptscriptfont\curly=\curlyfontss
\def\cal{\fam\curly\relax}
\fi
%

%\font\bbol=cmbx10 scaled \magstep1    %chapter titles
\ifx\boringfonts\fonttest\else\fi

%macros to get automatically-numbered figures
\global\newcount\pnfigno \global\pnfigno=1
\newwrite\ffile
\def\pfig#1#2{Fig.~\the\pnfigno\pnfig#1{#2}}
\def\pnfig#1#2{\xdef#1{Fig. \the\pnfigno}%
\ifnum\pnfigno=1\immediate\openout\ffile=figs.tmp\fi%
\immediate\write\ffile{\noexpand\item{\noexpand#1\ }#2}%
\global\advance\pnfigno by1}

% This one embeds figs in the text (don't mix \tfig with \pfig!)
% Unlike \pfig, this one comes in two parts: \tfig allocates the number and
% \ifig inserts the empty box
% e.g. blah, blah (\tfig\foo) blah, end of paragraph.
%
% \ifig\foo{caption.}{2.5}
%
% New paragraph; a box 2.5truein has been left and captioned.
\def\tfig#1{Fig.~\the\pnfigno\xdef#1{Fig.~\the\pnfigno}\global\advance\pnfigno
by1}

% here come Distler's versions
% Interface to epsf.tex
% the call (after assigning a figure number with \tfig) is
%
%\ifigure\figlabel{caption}{figfile}{vsize}
%
%If \figflag is undefined, it leaves a vbox with caption and .2truein
%of space (like \ifig).  If you have \def\figflag{y}, it inserts the figure
%scaled to fit. The logic here is that you can turn off all figure
%generation by commenting out one line at the beginning of the file
%(the line that \def's \figflag). The pagination, etc. is completely
%the same whether the figure is there or not.
%
%There is also a macro called \epsfsize, which checks that the figure is
%narrow enough to fit in the current \hsize. If not, it is again scaled to fit.
%
\def\figI{y}
\def\ifigure#1#2#3#4{
\midinsert
\ifx\figflag\figI
 \ifx\htflag\figI
 \vbox{
  \href{file:#3}% assumes figure has been loaded locally (no absolute
               % url capability in xhdvi); useful for a better look
{Click here for enlarged figure.}}
 \fi
 \vbox to #4truein{
 \vfil\centerline{\epsfysize=#4truein\epsfbox{#3}}}
\else
\vbox to .2truein{}
\fi
\narrower\narrower{\ninepoint\noindent{\bf #1:} #2}
\endinsert
}

% new stuff for bozo

%\def\hal{{1\over2}} %obsolete

%\def\sump{\CO({\textstyle{\sum_{i=1}^q}}Q_i-{\textstyle{\sum_{j=1}^q}}P_i)}

%%%%%%%%%%%%%% goodies for super riemann surfaces %%%%%%%%%%%%%%

 %beltrami diffl

\def\DD{\pd{}\theta+\theta\pd{}u}   % big D

\def\th#1#2#3{\hat{t}_{#1#2}^{\phantom{#1#2}#3}}

                     % bold z
                     % bold u
  % bold z-bar
  % bold u-bar

%
% A-hats
   %curly A-hat
   %curly A-hat, holo
   %curly A-hat, antiholo
%
% Omega-hats

%
% O-hats

%%%%%%%%%%%%%%%%%%%%%%%%%%%%%%%%%%%%%%%%

% Poor man's Blackboard Bold characters often used :
\def\inbar{\,\vrule height1.5ex width.4pt depth0pt}
\def\IB{\relax{\rm I\kern-.18em B}}
\def\IC{\relax\hbox{$\inbar\kern-.3em{\rm C}$}}
\def\ID{\relax{\rm I\kern-.18em D}}
\def\IE{\relax{\rm I\kern-.18em E}}
\def\IF{\relax{\rm I\kern-.18em F}}
\def\IG{\relax\hbox{$\inbar\kern-.3em{\rm G}$}}
\def\IH{\relax{\rm I\kern-.18em H}}
\def\II{\relax{\rm I\kern-.18em I}}
\def\IK{\relax{\rm I\kern-.18em K}}
\def\IL{\relax{\rm I\kern-.18em L}}
\def\IM{\relax{\rm I\kern-.18em M}}
\def\IN{\relax{\rm I\kern-.18em N}}
\def\IO{\relax\hbox{$\inbar\kern-.3em{\rm O}$}}
\def\IP{\relax{\rm I\kern-.18em P}}
\def\IQ{\relax\hbox{$\inbar\kern-.3em{\rm Q}$}}
\def\IR{\relax{\rm I\kern-.18em R}}
\font\cmss=cmss10 \font\cmsss=cmss10 at 10truept%!!! should be 7pt
\def\IZ{\relax\ifmmode\mathchoice
{\hbox{\cmss Z\kern-.4em Z}}{\hbox{\cmss Z\kern-.4em Z}}
{\lower.9pt\hbox{\cmsss Z\kern-.36em Z}}
{\lower1.2pt\hbox{\cmsss Z\kern-.36em Z}}\else{\cmss Z\kern-.4em Z}\fi}
\def\IGa{\relax\hbox{${\rm I}\kern-.18em\Gamma$}}
\def\IPi{\relax\hbox{${\rm I}\kern-.18em\Pi$}}
\def\ITh{\relax\hbox{$\inbar\kern-.3em\Theta$}}
\def\IOm{\relax\hbox{$\inbar\kern-3.00pt\Omega$}}

%%%%%%%%%%%%%%%%%%%%%%%%%%%%%%%%%%%%%%%%%%%%%%%%%%%%%%%%%%%%%%%%%%%%%%

\def\cmp#1{#1 }         %or remove for stuffy journals

% for electronic submission, double-space and figures at end: (also use boring
% fonts)
% (and insert harvmac with ``l'' disabled)

\long\def\suppress#1{}
\suppress{\def\boringfonts{y}  %for export

\baselineskip=20pt
\def\ifigure#1#2#3#4{\nfig\dumfig{#2}}
 %end listrefs

} %end suppress

% BEGINNING
\def\ifigure#1#2#3#4{
\midinsert
\ifx\figflag\figI
 \ifx\htflag\figI
 \vbox{
  \href{file:#3}% assumes figure has been loaded locally (no absolute
               % url capability in xhdvi); useful for a better look
{Click here for enlarged figure.}}
 \fi
 \vbox to #4truein{
 \vfil\centerline{\epsfysize=#4truein\epsfbox{#3}}
}
\else
\vbox to .2truein{}
\fi
{\baselineskip8pt\narrower\narrower\ninepoint\noindent{\bf #1:} #2\par}
\endinsert
}

\long\def\optional#1{}
\def\testp{T}

\Title{\vbox{\hbox{UPR--765T}
}}{\vbox{\centerline{Torsional Directed
Walks,  }
\vskip2pt\centerline{Entropic Elasticity, and DNA Twist Stiffness}
}}

\centerline{J. David Moroz and Philip Nelson}\smallskip
\centerline{Department of Physics and Astronomy, University of Pennsylvania}
\centerline{Philadelphia, PA 19104 USA}
\bigskip

\ifx\prlmode\testp
\noindent {\sl PACS: 02.40.-k, %Geometry, differential geometry and topology
87.22.Bt. % Membrane and subcellular physics and structure
87.15.-v, %  Molecular biophysics
87.10.+e, %  General, theoretical, and mathematical biophysics
87.15.By.%  Structure, bonding, conformation, configuration, and isomerism of biomolecules
}\fi
\ifx\answ\bigans \else\noblackbox\fi
\def\Avalue{49}\def\Aerror{}
\def\Cvalue{120}\def\Cerror{}
\def\Ceffvalue{98}
\def\Dvalue{50}\def\Derror{} % D=Q*78nm
\def\Nvalue{49}
\def\thresh{2}
\def\variance{0.013}
DNA and other biopolymers differ from classical polymers due to their
torsional stiffness. This property changes the statistical character
of their conformations under tension from a classical random walk to
a problem we call the ``torsional directed walk''. 
Motivated by a recent experiment on single
lambda-DNA molecules [Strick {\it et al.} Science
{\bf271} (1996) 1835], we formulate the torsional directed walk
problem and solve it {\it 
analytically} in the appropriate force regime.
Our technique affords a direct physical
determination of the microscopic twist stiffness $C$ and twist-stretch
coupling $D$ relevant for DNA functionality.
The theory quantitatively fits existing
experimental data for relative extension as a function of overtwist
over a wide range of applied force; fitting to the experimental data
yields the numerical values  $C=\Cvalue\Cerror\,$nm and
$D=\Dvalue\Derror\,$nm. Future experiments will refine these 
values. We also predict that the phenomenon of {\sl reduction of
effective twist stiffness by bend fluctuations} should be testable in
future single-molecule experiments, and we give its analytic form.

\Date{11 July  1997}\noblackbox
%\advance\pageno by1
%\draft

%%%%%%%%%%%%%%%%%%%%%%%%%%%%%%%%%%%%%%%%%%%%%%%%%%%%%%%%%%%%%%%%%%%%%%

% useful gadgets
\def\DD#1#2{{{\rm d} #1\over{\rm d} #2}}
\def\Lk{{\rm Lk}}\def\Wr{{\rm Wr}}
\def\kbt{k_{\rm B}T}

\def\crit{_{\rm crit}}
\def\ft{\tilde f}

\def\eff{_{\rm eff}}

\def\wo{\omega_0}
\def\nm{\,{\rm nm}}
\def\th{{\bf t}}\def\zh{{\bf z}}

\hfuzz=3truept

%\let\lref=\ref
%%%%%%%%%%%%%%%%%%%%%%%%%%%%%%%%%%%%%%%%%%%%%%%%%%%%%%%%%%%%%%%%%%%%%%
% Here come the references:

% LREFS
\lref\rSYSB{J. Shimada and H. Yamakawa, ``Ring-closure probabilities
for twisted wormlike chains,'' Macromolecules
{\bf17} (1994) 689--698;
D. Shore and R.L. Baldwin, ``Energetics of DNA twisting
I,'' J. Mol. Biol. {\bf170} (1983) 957--981.}
%\lref\rMNpromise{J.D. Moroz and P. Nelson, in preparation (1997).} 
\def\rMNpromise{(J.D. Moroz and P. Nelson, in preparation)}
\lref\rVoloknot{See for example A.V. Vologodskii and N.R. Cozzarelli,
      ``Conformational and thermodynamic properties of supercoiled DNA,''
    Annu. Rev. Biophys. Biomol. Struct, {\bf23} (1994) 609\pagin{--43}
and references therein.}
\lref\rFuller{F.B. Fuller, ``Decomposition of the linking number of a
closed ribbon,'' Proc. Natl. Acad. Sci. USA {\bf75} (1978)
3357\pagin{--3561}.} 
\lref\rSdC{A. Stasiak and E. Di\thinspace Capua, ``The helicity of DNA
in complexes with RecA protein,'' Nature {\bf299} (1982) 185.}
\lref\rRudnick{B. Fain, J. Rudnick, and S. Ostlund 
\cmp{``Conformations of linear DNA,''} Phys. Rev. {\bf E55} (1996)
7364\pagin{--7368}. }

\lref\rBenh{See for example C. Benham, \cmp{``Geometry and mechanics
of DNA superhelicity,''} Biopolymers {\bf22} (1983)
2477\pagin{--2495}.}

\lref\olddna{M.H.F. Wilkins, R.G. Gosling, and W.E. Seeds, Nature
{\bf167} (1951) 759.}
\lref\rCReviews{For reviews see 
M. Record, S. Mazur, P. Melancon, J. Roe, S. Shaner, and L. Unger,
\cmp{``Double helical DNA: conformations, physical properties, and
interactions with ligands,''} Annu. Rev. Biochem. {\bf50} (1981)
997\pagin{--1024}\optional{[p1009: C=23--53 cal
bp/(mol deg2); later results C=17; ESR C=40] };
P.J. Hagerman, ``Flexibility of DNA,"
Ann. Rev. Biophys. Biophys. Chem. {\bf17} (1988)
265\pagin{--286}\optional{ [3E-19]};
D.M. Crothers, J. Drak, J.D. Kahn, and S.D. Levene, ``DNA bending,
flexibility, and helical repeat by cyclization kinetics,''
Meth. Enzymology {\bf212} (1992) 3\pagin{--29}\optional{ [3.4E-19]}.}
\lref\rMRZ{D.P. Millar, R.J. Robbins, and A.H. Zewail, ``Time-resolved
spectroscopy of macromolecules,'' J. Chem. Phys. {\bf74} (1981)
4200\pagin{--4201}.\optional{[C=1.3E-19 erg cm=33nm in calf thymus dna]}} 
\lref\rTH{W.H. Taylor and P.J. Hagerman J. Mol. Biol. {\bf212} (1990)
363.\optional{[2.0E-19 erg cm=50nm]}}
\lref\rSCPWKH{P.R. Selvin, D.N. Cook, N. Pon, W.R. Bauer, M.P. Klein,
J.E. Hearst, ``Torsional rigidity of positively and negatively
supercoiled DNA,'' Science {\bf255} (1992)
82\pagin{--85}.\optional{[C=1.9E-19 erg cm=50nm]}}
\lref\rVALF{A.V. Vologodskii, V.V. Anshelevich, A.V. Lukashin, and
M.D. Frank-Kamenetskii, ``Statistical mechanics of supercoils and the
torsional stiffness of the DNA double helix,'' Nature {\bf280} (1979)
294\pagin{--298}.\optional{[(0.036RT/deg2)~.~0.34nm=165nm]}}
\lref\rfirst{S.B.~Smith, L.~Finzi and C.~Bustamante,
      \cmp{``Direct mechanical measurements of the elasticity of
single DNA molecules       by using magnetic beads,''}
    Science {\bf 258} (1992) 1122\pagin{--6};
C.~Bustamante, J.F.~Marko, E.D.~Siggia and S.~Smith,
     \cmp{``Entropic elasticity of lambda-phage DNA,''}
  Science {\bf 265} (1994) 1599\pagin{--600}.\optional{[early tweezer
measurements of A]}} 
\lref\rJMPRE{J.F.~Marko,  ``Supercoiled and braided DNA under tension,''
   Phys. Rev. {\bf E55} (1997) 1758\pagin{--72}.}
\lref\rMSc{J.F.~Marko
and E.D.~Siggia, ``Fluctuations and supercoiling of DNA,''
   Science {\bf265} (1994) 506\pagin{--508}.}
\lref\rMSb{J.F.~Marko
and E.D.~Siggia, ``Stretching DNA,'' Macromolecules {\bf 28} (1995)
8759\pagin{--8770}.}
\lref\rMSa{J.F.~Marko
and E.D.~Siggia, ``Bending and twisting elasticity of DNA,''
Macromolecules, {\bf 27} (1994) 981\pagin{--988}.}
\lref\rJMTS{J.F. Marko, ``Stretching must twist DNA,''
Europhys. Lett. {\bf 38} (1997) 183\pagin{--188}.} 
\lref\rLandau{L. Landau and E. Lifshitz, {\sl Theory of elasticity}
3$^{rd}$ ed.  (Pergamon, 1986).}
\lref\rMaVo{See for example J.F. Marko and A. Vologodskii, ``Extension
of torsionally stretched DNA by external force,'' Biophys. J. {\bf73}
(1997) 123-132.} 
\lref\rCalla{C. Calladine and H. Drew, {\sl Understanding DNA} (Academic,
1992).}
\lref\roverstretch{P. Cluzel, A. Lebrun, C. Heller, R. Lavery,
J.-L. Viovy, D. Chatenay, and F. Caron, \cmp{``DNA: an extensible
molecule,''} Science {\bf271} (1996) 792\pagin{--794}; S. Smith, Y. Cui, and
C. Bustamante, \cmp{``Overstretching B-DNA: the elastic response of
individual double-stranded and single-stranded DNA molecules,''}
Science {\bf271} (1996) 795\pagin{--799}\optional{[12/15/95]}.}
\lref\rWang{M.D. Wang, H. Yin, R. Landick, J. Gelles,
and S.M. Block,
\cmp{``Stretching DNA with optical tweezers,''}
Biophys. J. {\bf72} (1997) 1335\pagin{--1346}.}
\lref\rStrick{T.R. Strick, J.-F. Allemand, D. Bensimon, A. Bensimon, and V.
Croquette, \cmp{``The elasticity of a single supercoiled DNA
molecule,''}
Science {\bf271} (1996) 1835\pagin{--1837}.}

\lref\rCDKL{D.M. Crothers, J. Drak, J.D. Kahn, and S.D. Levene, ``DNA
bending,'' Meth. Enzymology {\bf212} (1992)3\pagin{--29}.}

\lref\rHag{P.J. Hagerman, ``Flexibility of DNA,"
Ann. Rev. Biophys. Biophys. Chem. {\bf17} (1988) 265\pagin{--286}.}
\lref\rKLNO{R.D. Kamien, T.C. Lubensky, P. Nelson,
and C.S. O'Hern, ``Direct Determination of DNA Twist-Stretch Coupling,''
Europhys. Lett. {\bf38} (1997) 237\pagin{--242}; R.D. Kamien, T.C. Lubensky, P. Nelson,
and C.S. O'Hern, ``Elasticity Theory of a
Twisted Stack of Plates'' preprint 1997 [cond-mat/9707040].} 
\lref\rKM{The scaling limit of
the torsional random walk has been studied recently in J.D. Moroz and R.D. Kamien,
``Self-avoiding walks with 
writhe,'' preprint 1997 [cond-mat/9705066].}
\lref\rVa{A.V. Vologodskii, ``DNA extension
under the action of an external force,'' Macromolecules {\bf27} (1994)
5623\pagin{--5625}.}

\lref\rBM{C. Bouchiat and M. M\'ezard, ``Elasticity model of a supercoiled
DNA molecule,'' preprint 1997 [cond-mat/9706050].}
\lref\rLove{A.E.H. Love, {\sl A treatise on the mathematical theory of
elasticity} 4th   ed. (Dover, 1944), \S272d.}
\lref\rDoiE{M. Doi and S.F. Edwards, {\sl The theory of polymer
dynamics,} (Oxford, 1988).}
\lref\rAustin{See R.H. Austin, J.P. Brody, E.C. Cox, T. Duke, and
W. Volkmuth, ``Stretch genes,'' Physics 
Today {\bf50} (1997) 32\pagin{--38} and references therein.}

%%%%%%%%%%%%%%%%%%%%%%%%%%%%%%%%%%%%%%%%%%%%%%%%%%%%%%%%%%%%%%%%%%%%%%
% BEGINNING
\newsec{Introduction and Summary}
The theory of random walks is one of the most fundamental problems in
statistical mechanics, with applications throughout physics, biology,
and even finance. The discovery that polymer conformations
afford a concrete realization of this mathematical problem, and the
understanding that rubber elasticity is inherently an {\it entropic}
phenomenon, marked the birth of polymer physics \rDoiE. Remarkably, it
has recently become possible to apply minuscule forces to
{\it single molecules} of DNA in solution and observe their
extension~\rAustin. Besides allowing a detailed confirmation of the
directed random walk model of entropic elasticity, these experiments
allow direct physical measurement of microscopic (nanometer-scale)
mechanical properties of DNA relevant to its function, using
mesoscopic (micron-scale) apparatus. Two linear elastic
parameters of DNA have now been measured in this way: the bend
persistence length $A$ and the intrinsic-stretch modulus $\ga$
\rfirst\rVa\rMSb
\roverstretch
\rWang.

DNA and other stiff biopolymers differ from classical polymers,
however, in that they exhibit {\it torsional} as well as bend
stiffness. Thus their conformations reflect not a classical directed
walk but a new fundamental problem: the ``torsional directed walk'' (TDW),
whose random variables are the direction of each step relative to its
predecessor, together with a relative axial twist. In this letter we
will formulate the version of the TDW appropriate to DNA,
solve it {\it analytically} in a regime appropriate to a recent experiment
\rStrick, and show that the model quantitatively fits the data over a
wide range of applied forces (\tfig\fone) \rKM. Besides being
transparent, analytic formul\ae\ facilitate systematic least-squares
fitting to experimental data. 
%Our calculation uses a continuum model with no artificial
%short-length cutoff. 
We fit to obtain three
microscopic elastic constants: the twist persistence length $C$, bend
persistence length $A$, and intrinsic twist-stretch coupling
$D$ \rJMTS\rKLNO. Since $A$ is known independently we have a check on the
model. The experiment is not sensitive to the  other allowed
linear-elastic constants such as twist-bend coupling \rMSa.

\ifigure\fone{Relative extension of lambda-DNA versus applied force $f$
and overtwist $\sigma$. From top to bottom, the curves are at fixed
force 8.0, 1.3, 0.8, 0.6, 0.3, and 0.1~pN. The dots are experimental
data from Fig.~3 of \rStrick, excluding values of $f,\sigma$ where the DNA is
known to denature by strand separation. Open dots are outside the range of
validity of the phantom chain model and were not used in the
fit. The corresponding points from Fig.~2 of \rStrick\ were also used
in the fit (not shown), for a total of \Nvalue\ points. The lines are our
theoretical predictions 
%after fitting to $A,C$, and $D$ (see text).}{fig1kaleida.eps}{4.5}
after fitting to $A,C$, and $D$ (see text).}{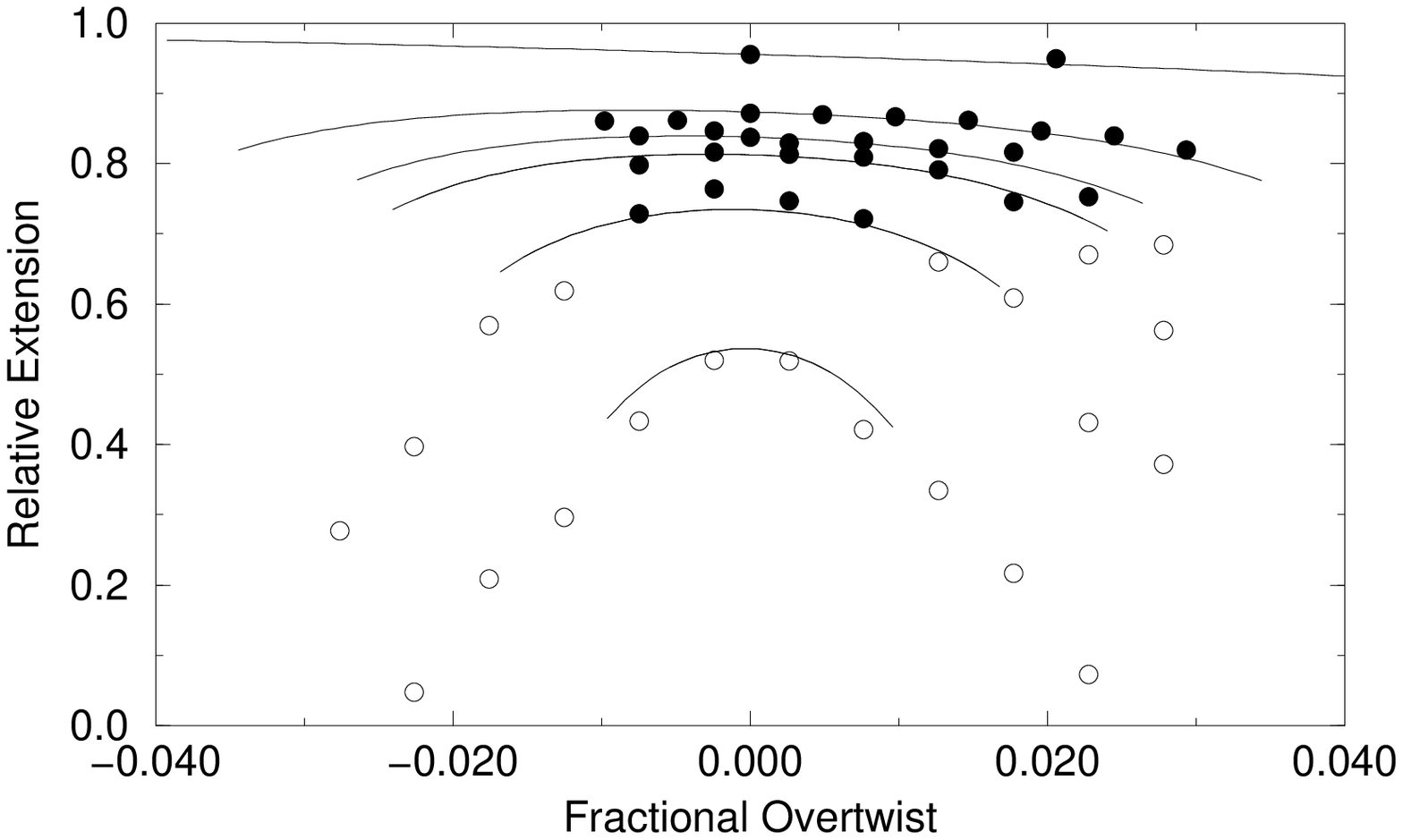}{4}

We find that the existing data \rStrick\ yield 
$A=\Avalue\Aerror\,$nm, $C=\Cvalue\Cerror\,$nm, and  $D=\Dvalue\Derror\,$nm;
future experiments will refine these values when fit to our
formula. Many authors have sought to extract the value of $C$ from
both cyclization experiments and fluorescence depolarization
\rCReviews 
. A key point of this paper is that
the force regime we study is free from some vexing physical and mathematical
difficulties which have helped make the determination of $C$ from
these experiments
controversial. In particular,
we can use a continuum model with no need for the
short-length cutoff required to make Monte Carlo calculations
tractable \rMaVo. 

We also give a simple analytical prediction for the {\sl reduction of
effective twist stiffness by bend fluctuations.} This renormalization
may explain why some other determinations of $C$ give slightly lower
values than ours. While its existence was appreciated long ago by Shimada
and Yamakawa \rSYSB, a clear experimental test has hitherto not been possible.
While this effect is only marginally visible in the
extant data, again future experiments should be able to test our
prediction by checking the dependence of $C\eff$ on the applied force
(see below).  Our value for $D$ is similar to within the large errors
to recent estimates \rJMTS\rKLNO.

In a recent preprint Bouchiat and M\'ezard have independently
addressed several overlapping issues \rBM. We comment on their
approach below.

\newsec{Experiment}
In the experiment of Ref.~\rStrick, DNA was bound to a wall at one end
and a magnetic bead at the other, with bonds which did not permit free
pivoting. Constraining the orientation of the bead with an applied
magnetic field thus constrained the orientation of the DNA strand at
its end. Since the bead was too large for the DNA to loop around it,
this procedure effectively fixed the {\it total Link} of the circuit
consisting of the DNA plus a fixed imaginary closing path. Rotating
the applied magnetic field then allowed the authors to freeze the Link
to any desired value and find the extension $Z$ for various values
of the applied stretching force $f$. Alternatively, the force could be
held fixed while the Link was 
varied, as in \fone\ above. The intensive strain variable describing
Link is the {\it relative overtwist} $\sigma\equiv\Delta{\rm
Lk}\cdot(3.4\,{\rm nm}/L)$, where $L=16400\,$nm is the total contour length.

Subsequent work showed that the DNA
denatures by strand separation for high applied stresses, roughly
$f>0.4\,$pN and $\sigma<-0.01$ or $\sigma>0.03$
(D. Bensimon, private communication). The loss of twist
rigidity is also clearly visible in the curves in \rStrick. We have 
omitted such
points from \fone. Also, DNA undergoes a dramatic overstretching
transition at around 60~pN \roverstretch; all the data discussed here
concern forces $f\le 8\,$pN.

\newsec{Physical Picture}
We will begin with a picture of DNA as a thin cylindrical elastic rod
of fixed contour length $L$;
below we will discuss corrections reflecting the more detailed
architecture of the molecule. The conformations of such a rod under an
applied tension are
controlled by the elastic energy functional\foot{Our notation is
similar to \rMSa. Throughout we
will neglect sequence dependence. 
In the force regime in question we expect linear rod elasticity to be
a good approximation. Higher-order bend
elasticity effects are expected to be suppressed by powers of the rod
radius (1~nm), which is much smaller than any other length scale.
Indeed a linear-elastic model, the ``extensible worm-like
chain'', describes accurately the extension of torsionally
unconstrained DNA up to forces greater than those considered here,
with an intrinsic stretch modulus more than a hundred times 
greater than the forces of interest to us~\rWang. } 
\eqn\eroda{E/\kbt=\half\int\left[
A((\Omega_1)^2+(\Omega_2)^2)+C(\Omega_3-\omega_0)^2
\right]\dd s - \ft\cdot Z-2\pi\tau\cdot\Lk\quad.}
Here $s$ is arc length, $\Omega_{1,2}(s)$ are bending strains,
$\Omega_3-\wo$ is twist strain, $A,C$ are the bend and twist
persistence lengths, $\ft\equiv 
f/\kbt$, and $\wo\equiv2\pi/3.4\nm$ is the unstressed molecule's helix
density.  $\tau$ is a dimensionless torque variable which we will
choose to obtain the required overtwist $\langle\Lk\rangle=(1+\sigma)L\wo/2\pi$.

In the absence of thermal fluctuations, a rod under extensional stress
remains straight as we apply increasing torque to the ends, then
becomes linearly unstable at a critical value of torque which increases
with the applied force~\rLove.  The end-to-end distance $Z$ of the
rod does not change at all for applied overtwist less than the
critical value $\tau\crit=2\sqrt{A\ft}$, since the rod remains a
straight line of constant contour 
length. 

Thermal fluctuations change this picture completely. The rod is never
straight; every Fourier mode of its shape is excited in accordance
with the equipartition theorem of statistical physics, so the net
length $Z$ is always less than $L$. The applied tension suppresses
those modes of wavenumber smaller than $q_0\equiv\sqrt{\ft/A}$, with
the dominant contribution to $1-(Z/L)$ coming from $q\approx q_0$.
Now when we apply torque below the
critical value, the
modes with the same helical sense as the torque get pushed {\it closer
to} instability and their fluctuations are enhanced, shortening the
rod. The modes with the opposite helical sense are suppressed, and to
first order in $\sigma$ there is no effect, consistent with the
obvious fact that the elastic energy \eroda\ does not break inversion
symmetry. We {\it do} expect an effect to $\CO(\sigma^2)$; indeed this is the
dominant feature of the data in \fone. As the force decreases the
effect increases, as seen in the increasing curvature of the curves in
the figure. 

In eqn.~\eroda\ we have neglected any self-avoidance effects; these
would appear as interactions between rod elements distant in $s$. In
the usual directed walk this is not a serious omission: the crossover
to self-avoiding-walk scaling occurs only for chain lengths much
longer, and forces much smaller, than those encountered in DNA. In the
TDW we must be more careful, since the linking number appearing in
\eroda\ is undefined when the chain crosses itself. Physically the
problem is that the chain can form a loop, pass through itself, and in
the process lose a unit of Link: the {\it phantom torsional chain
cannot support any imposed torque.} If we seek an equilibrium at
nonzero $\tau$ we must expect to find mathematical pathologies; they
will arrive in due course. A related problem is that the statistical
sum for the phantom torsional chain includes all {\it knotted}
configurations, an error with noticeable effects \rVoloknot.

One approach to this problem is to introduce realistic self-avoidance
and knot rejection into eqn.~\eroda\ \rVoloknot\rMaVo. The resulting
nonlocal model requires numerical Monte Carlo solution, and the
results depend on the details of the chain interaction chosen. From
the physical picture, however, it is clear that at high enough applied
tension $f$ the problematic loops and knots will be so rare as to be
negligible: the chain remains nearly straight, and we may use \eroda\
without modification. We will find below the precise condition to be
in this regime; it corresponds to the solid dots in \fone.

Eqn.~\eroda\ also neglects any effects of {\it rod anisotropy}. For
example, bending into the major groove (``Roll'') is obviously easier
than bending in the perpendicular direction (``Tilt''); less obvious
is an allowed twist-bend coupling \rMSa. Such anisotropies can lead to
chiral effects, for example an asymmetry between $\sigma$ and
$-\sigma$, but their effects on \fone\ are negligible
\rMNpromise. Indeed we only expect the helical pitch to affect
entropic elasticity when the dominant wavenumber $q_0$ approaches
$\wo$, \ie\ at unattainably large forces.\foot{Certainly the omitted
anisotropic couplings will also renormalize the constants $A,C$ in
\eroda\ \rMNpromise. The model \eroda\ is to be regarded as
coarse-grained to the scale of the helix pitch.}

Instead the asymmetry visible in \fone\ has its origin in the {\it
intrinsic stretch} elasticity of a chiral rod, which gives $Z$ a
contribution linear in $\sigma$ and independent of $f$ \rKLNO. This
coupling may be relevant for the binding of the protein RecA to DNA,
which 
stretches and untwists the DNA \rSdC. The effect should be masked by
the entropic $\sigma^2$ term for small force, emerging when the latter
is suppressed at high force, exactly as seen in \fone.

\newsec{Calculation}
We now sketch a calculation embodying the above physical picture
\rMNpromise. We introduce three local configuration variables: a unit
vector $\th(s)$ describing the tangent to the chain, and an angle
$\zeta(s)$ for the remaining torsional degree of freedom. We take the
applied force along the \zh\ direction and orient the chain so that
\th=\zh\ in equilibrium. To define $\zeta(s)$ we use Fuller's local
formula for the writhe of a curve whose tangent never points along the
-\zh\ axis \rFuller\rRudnick: Wr$=\lfr1{2\pi}\int\dd s(\th\times\DD\zh s)\cdot\zh/
(1+\th\cdot\zh)$. Combined with White's theorem that $2\pi\Wr+\int
\Omega_3\dd s$ is a topological invariant, we see that $\Omega_3+
(\th\times\DD\zh s)\cdot\zh/
(1+\th\cdot\zh)$ must be a total derivative; we will call this
quantity $\wo+\DD\zeta s$ and eliminate $\Omega_3(s)$ in favor of
$\zeta(s)$.\foot{A more elegant, equivalent, approach takes the
configuration variables to be a $3\times3$ rotation matrix; the torque
term then takes the form $-\tau\int\dd s[\Omega_3+\check\Omega_3]/(1
+\th\cdot\zh)$, where $\check\Omega_i$ are the {\it space-fixed}
angular velocities of a rigid body \rMNpromise.}
The advantage of this choice is that \eroda\ is now quadratic in
$\zeta$, which may be summarily eliminated.

It proves convenient to define dimensionless quantities
$K\equiv\sqrt{A\ft - \tau^2/4}$ and $\bar s\equiv Ks/A$; we then find 
\eqn\eEred{E/\kbt=\CE_0+\lfr K2\int\dd \bar s\left[
\|\dot\th\|^2+\lfr{2\ft A}{K^2}(1-\zh\cdot\th)
-\lfr{2\tau}K (\th\times\dd\th/\dd s)\cdot\zh/
(1+\th\cdot\zh)
\right]\ ,}
where $\CE_0\equiv-L(\ft+\tau^2/2C)$ and dot denotes $\DD{}{\bar
s}$. The second term defines a nonlinear fluctuation problem, which we
will expand in powers of $1/K$. From its partition function
$\CZ(f,\tau)$ we may then extract the extension and excess link as
\eqn\edlogs{Z=\left.\pd{}{\ft}\right|_\tau\log \CZ\ ,\quad
\wo(1+\sigma)={1\over L}\left.\pd{}\tau\right|_{\ft}\log\CZ\ .}
We will use the second of these to solve for $\tau(f,\sigma)$, then
substitute into the first to get the desired extension $Z(f,\sigma)$.

To find $\CZ$ we adapt the standard trick used in the wormlike chain
\rDoiE\rMSb: for long chains \CCZ\ approaches the unnormalized
correlation function of \th. Holding $\th(0)$ fixed, this correlator
$\psi(\th,s)$ obeys the Schr\"odinger-like equation $\dot\psi=-H\psi$,
where \rMNpromise
\eqn\eH{H=K(1-\cos\theta)-\lfr1{2K}{\bf L}^2
+\lfr\tau K(1+\cos\theta)\inv\,L_z 
+\lfr{\tau^2}{2K}\bigl(\half-(1+\cos\theta)\inv\bigr)(1-\cos\theta)\ .
}
Here $\cos\theta=\th\cdot\zh$, ${\bf L}^2$ is the angular part of the
Laplace operator, and $L_z$ is the azimuthal derivative. The main
novelty of this derivation is the presence of first-order derivatives
in \eEred\ when $\tau\not=0$, leading to the $\tau^2$ terms in $K$ and in
\eH. The asymptotic value of \CCZ\ is then controlled by the lowest
(``ground-state'') eigenvalue of \eH\ via $\CZ\propto\ex{-\CE_0-
\lambda_0L}$.

Unfortunately \eH\ has {\it no} ground state for any nonzero $\tau$, due to
its singularity at $\theta=\pi$! This unphysical pathology was
predicted in the previous section; its mathematical origin is the
breakdown of Fuller's formula when $\th=-\zh$. To see the connection
to the physical discussion, note that the unphysical link-dropping
process in the phantom torsional chain necessarily involves the
tangent $\th(s)$ passing through -\zh\ at some intermediate point. As
discussed above, a physically meaningful and analytically tractable
resolution to the problem is to restrict attention to large $f$. We
can then solve \eH\ in perturbtion theory about $\theta=0$, where the
problem is invisible, provided the perturbative ground state value is
smaller than the ``tunneling barrier'' of \eH.\foot{To justify
perturbation theory itself we note that it gives an excellent
approximation to the exact solution of the wormlike chain \rMSb\ when
$K>1$, as may be expected from the form of the leading anharmonic
correction below. Nonperturbative effects in the variational approach
to the WLC \rMSb\ are also small when $K>1$. Note that our data cuts
also eliminate the region $\sigma>2\pi/C\wo$ where plectoneme
formation (and hence large self-avoidance effects) is expected
\rJMPRE. Raising our threshold on $K$ selected fewer points with
little effect on our result.
} Imposing this condition and $K^2>\thresh$ selects the solid dots in
\fone. 

The perturbative solution of \eH\ gives the ground
state eigenvalue  $\la_0=1-1/4K-1/64K^2\cdots$, so \edlogs\ gives 
$\sigma=\lfr\tau\wo\bigl(C\inv+(4KA)\inv+\CO(K^{-3})\bigr)$,
$Z/L=1-(2K)\inv(1+1/64K^2+\CO(K^{-3})$. Solving by iteration gives the
torque
\eqn\eCren{\tau(f,\sigma)=\wo\sigma/\Bigl[C\inv+\bigl(4A\sqrt{Af/\kbt}\bigr)
\inv\Bigr]\ ,}
plus corrections of $\CO(K^{-3})$. Formula \eCren\ displays the
promised renormalization of twist stiffness by bend
fluctuations. While direct torque measurements are not currently
possible, this effect nevertheless enters the force curves since
$\tau$ enters $Z/L$.

Assembling the pieces gives our theoretical prediction for the force
curve: the relative extension $Z/L$ is
\eqn\efinal{Z(f,\sigma)/L=1-\half(Af/\kbt - (\tau/2)^2-1/32)^{-1/2}
+(f-D\sigma)/\gamma+A/LK^2\ ,}
with $\tau$ from \eCren. We have improved the previous formula
\rMNpromise\ by summing perturbation theory,  introducing the
intrinsic-stretch terms mentioned in the previous section, and
including a small finite-length correction.

In formulas \eCren\efinal\ the parameters $\wo,L$ are known and we use
the value $\gamma=1100\,$pN for the linear stretch constant obtained
from higher-force experiments \rWang. This leaves $A,C$, and $D$ which
we fit to the experimental data after the cuts described above. \fone\
shows that a single choice of $A,C,D$ fits all the curves.%
\foot{The variance of the data from our curves is $\sigma_{Z/L}=\variance$,
comparable 
to the visible scatter in the data. The formal covariances for $A,C,Q$
correspond to very small errors; in practice the fit is visibly worse
for $C$ outside the range $70<C<150$.}
A
nontrivial check is that several points just outside our accepted set,
not used in the fit, nevertheless lie on our theoretical curves.

Bouchiat and M\'ezard have recently analyzed the $f=0.1\,$pN curve
\rBM. They independently obtained formul\ae\ equivalent to
\edlogs\eH. Since the 0.1~pN curve is  entirely outside the
range of validity of the phantom chain model, however, they were
obliged to address the unphysical pathology of \eH, both by Monte
Carlo simulation and by introducing a new short-scale cutoff. The new
cutoff introduces a new unknown parameter into the theory, and
moreover does not correspond in a simple way to the actual physics of
self-avoidance. Nevertheless they found an impressive fit to the
0.1~pN curve and relative insensitivity to their choice of cutoff for
$|\sigma|<0.015$.

Formulas \eCren\efinal \ make many predictions. We note that the fit
value of $A$ is consistent with \rWang. Omitting either the $f/\gamma$
or the $-D\sigma/\gamma$ terms makes a poorer fit, as does
replacing \eCren\ by the na\"\i ve $\tau=\wo\sigma/C\eff$ 
with constant $C\eff$. 
(It 
is interesting to note that the value $C\eff=\Ceffvalue\,$nm obtained in this way
is closer to earlier determinations than our microscopic value of $C$.)
Definitive test of \eCren\ must await further experiments.

\ifx\scimode\testp
\vskip .2truein\noindent{\sl Correspondence to P. Nelson.}
\fi

\ifx\scimode\testp
\footatend\bigskip\immediate\closeout\rfile\writestoppt
  \baselineskip=22pt\centerline{{\bf References}}\bigskip{\frenchspacing%
  \parindent=20pt\escapechar=` \input refs.tmp\vfill\eject}\nonfrenchspacing
 \vfill\eject\immediate\closeout\ffile{\parindent40pt
 \baselineskip22pt\centerline{{\bf Figure Captions}}\nobreak\medskip
 \escapechar=` \input figs.tmp \vfill\eject
}

\vskip .5truein
{\frenchspacing
We would like to thank B. Fain, 
R.D. Kamien, T.C. Lubensky, C. O'Hern, and J. Rudnick
for helpful discussions, and D. Bensimon and
V. Croquette for supplying us with experimental details and the
numerical data from \rStrick. This work 
was supported in part by  NSF.  JDM 
was supported in part by an FCAR Graduate Fellowship from 
the government of Quebec.}
\else

\vskip .5truein
{\frenchspacing
We would like to thank B. Fain, 
R.D. Kamien, T.C. Lubensky, C. O'Hern, J. Rudnick, and M. Zapotocky
for helpful discussions, C. Bouchiat and M. M\'ezard for
correspondence, and D. Bensimon and 
V. Croquette for supplying us with experimental details and the
numerical data from \rStrick. This work 
was supported in part by  NSF grant DMR95--07366.  JDM 
was supported in part by an FCAR Graduate Fellowship from 
the government of Quebec.}
\footatend\bigskip\immediate\closeout\rfile\writestoppt
  \baselineskip=22pt\centerline{{\bf References}}\bigskip{\frenchspacing%
  \parindent=20pt\escapechar=` \input refs.tmp\vfill\eject}\nonfrenchspacing
 \vfill\eject\immediate\closeout\ffile{\parindent40pt
 \baselineskip22pt\centerline{{\bf Figure Captions}}\nobreak\medskip
 \escapechar=` \input figs.tmp \vfill\eject
}

\fi
% \vfill\eject\immediate\closeout\ffile\parindent40pt
% \baselineskip22pt\centerline{{\bf Figure Captions}}\nobreak\medskip
% \escapechar=` \input figs.tmp \vfill\eject

%\else  %\scimode

\bye